\documentclass{svjour2}                    
\smartqed  
\usepackage{graphicx}
\usepackage{textcomp}
\begin{document}

\title{Study of superfluid $^3$He under nanoscale confinement
}

\subtitle{A new approach to the investigation of superfluid $^3$He films}


\author{L. V. Levitin\and R. G. Bennett\and A. Casey\and B. Cowan\and J. Saunders\and D. Drung\and Th. Schurig\and J. M. Parpia\and B. Ilic\and N. Zhelev} 


\institute{L. V. Levitin, R. G. Bennett, A. Casey, B. Cowan, J. Saunders\at
              Department of Physics, Royal Holloway University of London, Egham, Surrey, TW20 0EX, United Kingdom \\
              Tel.: +44 )0)1784 443486\\
              Fax: +44 (0)1784 472794\\
              \email{j.saunders@rhul.ac.uk}           
           \and
           D. Drung, Th. Schurig\at
              Physikalisch-Technische Bundesanstalt, Abbestrasse 2-12, D-10587 Berlin, Germany
\and J. M. Parpia, R. Ilic, N. Zhelev\at Department of Physics, Cornell, University, Ithaca, New York 14853, USA}

\date{Received: date / Accepted: date}

\maketitle

\begin{abstract}We review recent experiments in which superfluid $^3$He has been studied under highly controlled confinement in nanofluidic sample chambers. We discuss the experimental challenges and their resolution. These methods open the way to a systematic investigation of the superfluidity of $^3$He films, and the surface and edge excitations of topological superfluids.
\keywords{Superfluid $^3$He \and Film \and Topological superfluid}
\end{abstract}

\section{Introduction}
\label{intro}
As has been emphasized by Volovik \cite{Volovik}, superfluid $^3$He is a unique condensed matter system. There are two distinct superfluid phases: the time reversal invariant B-phase, and the chiral A-phase, which breaks time reversal symmetry. The phenomena predicted to emerge from these quantum vacua encompass a broad spectrum of modern physics. There is however limited experimental information on the surface and edge states which are predicted to appear through bulk-surface(edge) correspondence as a consequence of topologically non-trivial order parameter\cite{HasanKane,QiZhang}. Furthermore, despite the discovery of these p-wave superfluids in 1972 \cite{Leggett,Vollhardt}, there is little understanding of superfluid $^3$He \emph{films}, and the influence of film-thickness on the order parameter.

Our major experimental advance, reviewed here, is the ability to perform nuclear magnetic resonance on a single superfluid film of well defined thickness \cite{Levitin}. This initiates a new direction of quantum fluids research, combining state-of-the-art techniques in SQUID NMR (detection of the Nuclear Magnetic Resonance free precession signal using Superconducting Quantum Interference Devices), nanofabrication, and ultra-low temperature methods, in order to study the superfluidity of nanofluidic samples of $^3$He. It is founded on a series of technical breakthroughs: to confine $^3$He in a nanofabricated rectangular cavity of thickness comparable to the coherence length; to perform a precise characterisation of the geometry and surfaces of cavity; to cool the $^3$He sample to well below the superfluid transition temperature; to develop an NMR spectrometer of unprecedented sensitivity, which allows us to fingerprint the superfluid order parameter of these small samples. These developments provide the foundation for research on topological superfluids under nano-scale confinement, of importance in the rapidly developing field of topological quantum matter.

This  paper is organized as follows. We first introduce the superfluid $^3$He film, briefly reviewing previous experimental work, and set the context for different experimental approaches to the study of topological superfluid $^3$He surfaces. We summarise the technical advances made in our work, and briefly review the results obtained so far on superfluid $^3$He films. We conclude with an outline of key questions and future prospects.  

\section{Superfluid $^3$He film}
\label{sec:SF}
The discovery of superfluidity of a $^4$He film followed shortly after the discovery of superfluidity in the bulk liquid. The Rollin film, in the famous emptying beaker experiment \cite{Daunt}, is a saturated film. It resides on the container surface above the bulk liquid surface, and in its superfluid state can flow without resistance. The film thickness
as a function of height above the liquid surface is determined by the competition between the van de Waals attraction of helium to the container wall and gravity. Much later it was discovered \cite{Bishop} that superfluidity persists in unsaturated atomically thick films, and that the superfluid transition in such films is governed by the unbinding of vortex-antivortex pairs according to the Berezinskii-Kosterlitz-Thouless mechanism. This transition is a paradigm for defect-mediated phase transitions in two dimensions. Superfluidity survives in atomically thin films on disordered surfaces, above some threshold coverage necessary to smooth the surface disorder potential, since the superfluid coherence length is on the atomic length scale.

By contrast the coherence length in superfluid $^3$He has a lower bound given by the diameter of the p-wave pairs, $\xi_0=\hbar v_F/2 \pi k_B T_c$, where $k_B$
is Boltzmann's constant, $\hbar$ is Plank's constant divided by $2 \pi$, $v_F$ is the Fermi velocity and $T_c$ is the superfluid transition temperature. It diverges approaching the transition temperature $T_c$. Unconventional superconductor/superfluids are sensitive to disorder/scattering on length scales smaller than the coherence length, so the superfluid film needs to be thicker than roughly the pair diameter to avoid complete suppression of superfluidity, on a randomly scattering (diffuse) surface \cite{Kjaldman}. Surfaces with atomic scale roughness will scatter $^3$He quasiparticles diffusely. Studies of such thick films are complicated by instabilities in film morphology due to surface tension and a lack of precise knowledge of the key parameter: the thickness of the film. We do not address the potential superfluidity of atomically thin films of $^3$He; this will be influenced by both the dimensionality of the normal liquid and the surface disorder.

The superfluidity of the $^3$He film has been demonstrated in a series of film flow experiments: the classic emptying beaker experiment \cite{Steel}; a torsional oscillator measurement of the superfluid density \cite{Xu}; the detection of third sound \cite{Schechter}; and electrostatically driven and detected film flow \cite{SeamusDavis,Saitoh}. In these cases the surfaces are polished metal and not well characterised. Since the zero temperature coherence length is 70 nm, films need to be relatively thick on a diffuse surface to avoid suppression of superfluidity. 
While there are signs of some intriguing anomalies, indicative of potential new phases as the film is thinned, these flow experiments provide no direct insight into the superfluid order parameter. 

An alternative approach has been to confine the superfluid $^3$He in a stack of typically 1000 (thin mylar) sheets immersed in bulk liquid; the sheets are separated by some spacer material to define the film thickness. In these experiments the film thickness is fixed, but the effective confinement can be varied by exploiting the pressure dependence of $\xi_0$. Here the disadvantages include a significant distribution in spacing and the need to separate the confined signal from bulk. In the first such experiments both the superfluid density and the NMR response were investigated \cite{Freeman88,Freeman90} for 300 nm spacing; the equilibrium state was found to be A phase at all pressures. Later NMR studies used thicker nominal spacings of 0.8 \cite{Kawasaki}and 1.1~\textmu{}m \cite{Miyawaki}. In this case the film is expected to undergo an A to B phase transition, however the experimental evidence of the B phase and of this transition was unclear highlighting the importance of cavity thickness uniformity; we compare these results with our subsequent work in \cite{Levitin}. 

This earlier work made clear the requirement to study $^3$He within a well defined slab-like cavity, using modern methods of nanofabrication to define the sample morphology. It set a key goal: to identify the equilibrium order parameter as a function of film thickness.

\section{Topological superfluids; surface and edge excitations}
\label{sec:Topological}
New states of matter may be classified by their broken symmetry. A recent paradigm shift is the realization that a broad class of quantum condensed matter systems can be classified by their non-trivial topological order. Examples are the quantum hall effect \cite{Thouless}; the quantum spin hall effect \cite{KaneMele,Bernevig,Konig}; topological insulators \cite{FuKaneMele,MooreBalents,Hsieh,Xia}; topological superfluids and superconductors \cite{SalomaaVolovik,Schnyder,Roy,QiHughes,ChungZhang}. 

While investigations of candidate topological superconductors continue, there is renewed interest in the superfluid phases of $^3$He, which are firmly established topological superfluids. Superfluid $^3$He-B is a time reversal invariant odd-parity condensate of p-wave pairs, and the bulk is fully gapped, so the bulk density of quasiparticle excitations is exponentially suppressed below the transition temperature. The bulk topological invariant gives rise to gapless surface excitations through bulk-boundary correspondence. Quasiclassical theory \cite{Serene} provides a detailed framework to self consistently determine both the order parameter approaching the surface and the density of states of surface excitations, for different surface quasiparticle scattering conditions \cite{NagaiRough,NagaiAB,Vorontsov} . For specular scattering these surface excitations are Majorana fermions \cite{ChungZhang,Machida2010,WuSauls}, and the sample edge supports spin-current in the ground state. Superfluid $^3$He-A is a chiral superfluid; over the Fermi surface all the pairs have the same direction of their orbital angular momentum vector $\mathbf{l}$.  For specular boundaries there is no order parameter suppression at the surface, and no surface excitations. However distortion of the order parameter at the sample edge of a thin slab where the $\mathbf{l}$ is oriented uniformly across the slab in a single domain, leads to edge currents and an intrinsic orbital angular momentum \cite{Sauls2011}.

Thin film geometries engineered through nanofabrication, in which surface effects dominate, provide a new laboratory
of great potential for the investigation of these surface and edge effects. 

 An alternative strategy is to investigate the surface of the bulk superfluid. For example the nature of the coupling of positive ions to excitations at the free surface of superfluid $^3$He-B has been studied by ion mobility \cite{Ikegami}, a technique which has also been used to directly detect the chirality of the A-phase \cite{KonoScience}. Surface majorana excitations are helical, and approaches to detect this strong spin orbit coupling include studies of the spin dynamics of positive ions trapped below the helium surface \cite{Kono2013}, and studies of the spin-dynamics of
Q-balls \cite{Eltsov}. 

Another approach is to devise measurements to detect the mid-gap density of states of the surface excitations of the planar distorted B-phase at a solid surface. This will manifest itself in power law dependences of the heat capacity and thermal conductivity. The contribution of surface bound states to the heat capacity was already detected in silver sinters \cite{Choi}. An extensive series of measurements of the transverse acoustic impedance, at frequencies approaching the gap frequency, demonstrate clearly the existence of mid-gap surface excitations \cite{Murukawa}. Their density of states evolves, as expected theoretically, as the surface scattering condition is varied from diffuse toward specular. This technique allows the direct determination of specularity through measurements in the normal state. However it is clearly not compatible with measurements on fully specular surfaces. 

\section{Nanofluidic samples of superfluid $^3$He: Techniques}
\label{sec:NF}
\subsection{Nanofluidic cell}
\label{sec:NF1}

The first generation of sample cells were fabricated from glass (Hoya SD-2) and silicon wafers, anodically bonded together \cite{Dimov}. The unsupported cavity was defined lithographically in the silicon wafer, and characterised by a profilometer scan. The glass and silicon surfaces were characterised by AFM scans before bonding. The silicon surfaces are smooth with atomic scale roughness. The as-received glass wafers are potentially scratched, but a polishing procedure has been developed to eliminate these.

\begin{figure*}
\includegraphics[width=1.0\textwidth]{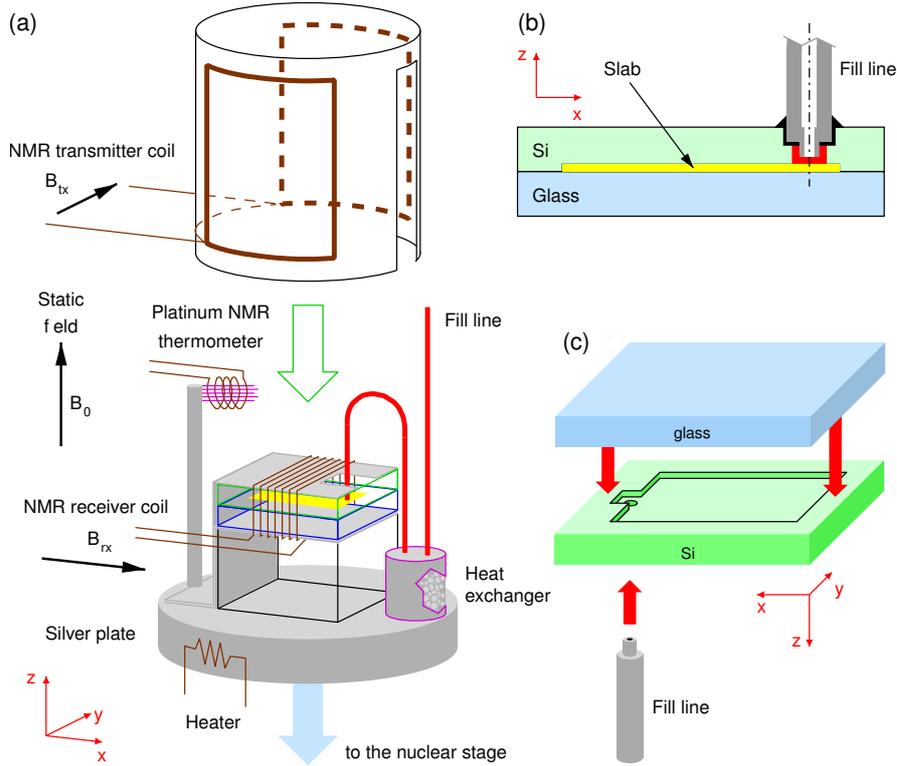}
\caption{Experimental apparatus. (a) The cell is mounted on the silver plate that also hosts the silver sinter heat exchanger where $^3$He is cooled and a Pt NMR thermometer.
The NMR receiver coil $L_{\mathrm{rx}}$ is tightly wound around the cell
and the NMR transmitter coil $L_{\mathrm{tx}}$ slides over.
(b) The nanofluidic cell with a slab of confined helium, yellow,
and a small bulk marker at the mouth of the fill line, red, visible to the NMR spectrometer.
(c) Essential steps of fabrication of the cell. A cavity is defined and the fill line hole is drilled in a silicon wafer.
A glass wafer is anodically bonded to it. Then the metallic fill line is glued in.}
\label{fig:setup}       
\end{figure*}

\begin{figure*}
  \includegraphics[width=0.75\textwidth]{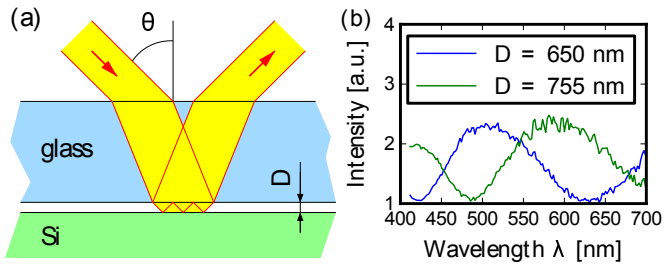}
\caption{Interferometry measurement of cavity thickness.
A collimated beam of white light of diameter 0.3~mm is reflected
off the cavity through the glass wafer (a).
The thickness is inferred from the observed interference between reflections
off the back of the glass wafer and the front of silicon wafer (b).}
\label{fig:optical}       
\end{figure*}

\begin{figure*}
  \includegraphics[width=1.0\textwidth]{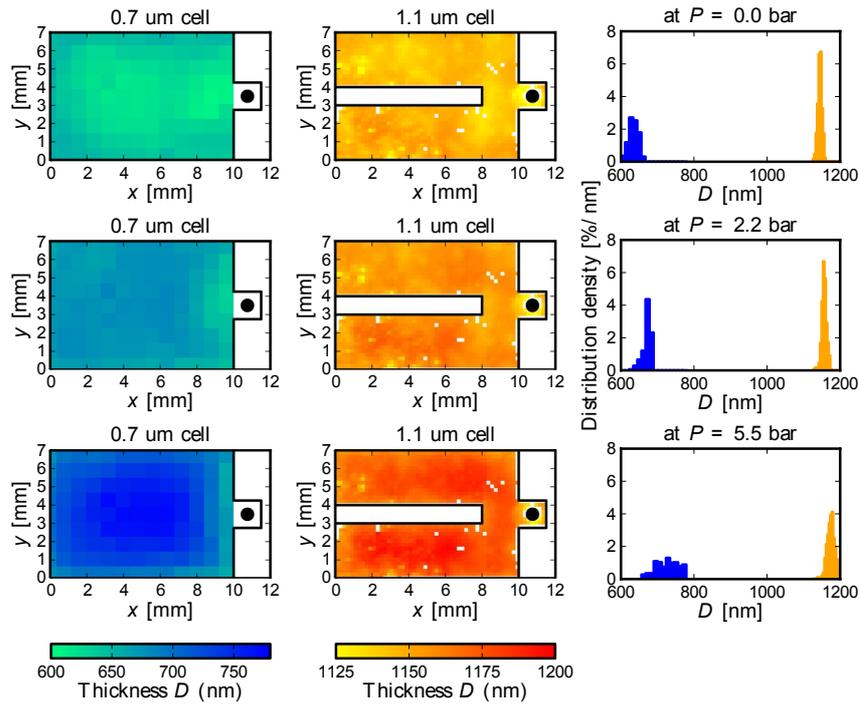}
\caption{Optical thickness measurements on 0.7 and 1.1~um cells at three different pressures
The 0.7~um cell shows bowing upon cooling due to differential thermal contraction
and inflation due to pressure which nearly cancel out at 2.2~bar. Both of these are reduced in the new 1.1~um cells by introducing a partition wall and refined bonding recipe at a lower temperature.}
\label{fig:thickness}       
\end{figure*}

The first cavity studied, of nominal unsupported height 650 nm, was fabricated from 3 mm thick wafers with a view to minimising changes in the cavity profile with $^3$He sample pressure, Fig.~\ref{fig:setup}b,c. The cavity profile as a function of pressure was determined by optical interferometry, through spectral analysis of a reflected collimated white light beam of diameter 0.3 mm, Fig.~\ref{fig:optical} . Cavity height profiles with this pixel size and a resolution of 2~nm are shown in Fig.~\ref{fig:thickness}. They indicate some bowing of the cavity due to differential thermal contraction, which stops at 30 K. The temperature independent deflection of the centre of the cavity under pressure is measured to be 28 nm/bar. The cavity distortion is accurately modelled by finite element analysis. A modified cavity, of nominal height 1050 nm has been fabricated with a central support and hence significantly smaller pressure dependent bowing. Optimized anodic bonding conditions have also improved the uniformity of cavity height.

\begin{figure*}
  \includegraphics[width=1.0\textwidth]{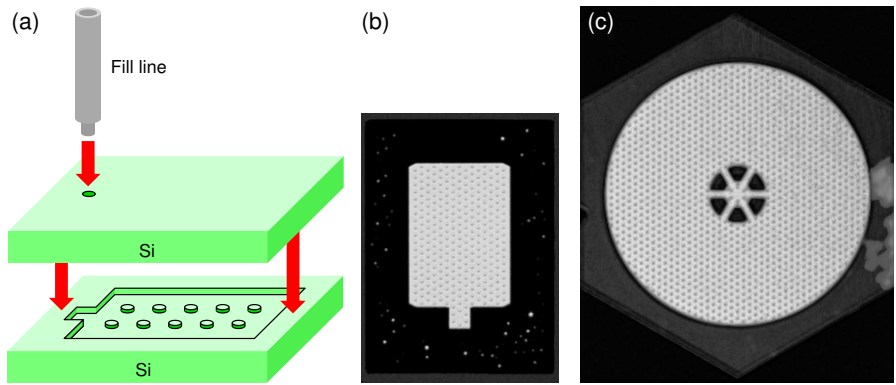}
\caption{Fully silicon cells with internal pillars allow to reduce the slab thickness and
open a new regime of confinement. Direct wafer bonding is used in fabrication.
Schematic fabrication diagram of an NMR cell (a).
The scanning acoustic microscopy scans of an NMR cell (b) and a torsional oscillator (c).}
\label{fig:SiCell}       
\end{figure*}

For thin cavities, the anodic bonding technique is not possible due to cavity collapse during bonding, and so we adopt a different method. We have fabricated a series of cells from two silicon wafers, with cavity heights 100, 200 and 300 nm. The two wafers are bonded by the direct wafer bonding technique. The cavity height is defined by a regular array of posts, patterned by optical lithography; here there is choice of the density of the array. In the most recent designs the fill line hole is fabricated by deep reactive ion etching. Different cells for NMR characterization of the equilibrium order parameter (with rectangular shape) or for torsional oscillator measurements of the superfluid density (circular) have been fabricated, Fig.~\ref{fig:SiCell}. 

\subsection{ SQUID NMR of a $^3$He film}
\label{sec:NF2} 
\begin{figure*}
  \includegraphics[width=1.0\textwidth]{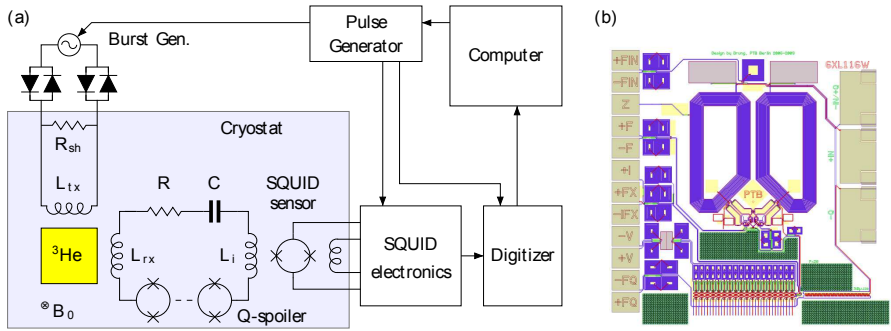}
\caption{Tuned dc~SQUID NMR spectrometer (a). The spins of $^3$He are tipped 
by a pulse in the transmitter coil $L_{\mathrm{tx}}$ and the free induction decay signal induced
in the receiver coil $L_{\mathrm{rx}}$ is detected
with a two-stage dc~SQUID sensor (b) shown as a single SQUID.
$L_{\mathrm{rx}}$ is coupled to the SQUID via a tank
circuit tuned at 1~MHz with quality factor $Q\sim30$.}
\label{fig:spectrometer}       
\end{figure*}

\begin{figure*}
  \includegraphics[width=0.75\textwidth]{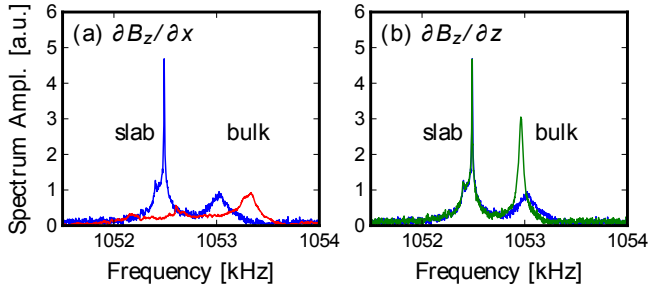}
\caption{Identification of cavity and bulk signals in the normal state (10~mK, 0~bar) \cite{Levitin}.
The NMR spectrum shows a double-peak structure.
By performing NMR in a field with externally applied gradients $\partial B_z/\partial x$
in the plane of the cavity and $\partial B_z/\partial z$ normal to it
we infer that the left- and right-hand peaks originate from the slab of confined helium
and a volume of bulk liquid at the mouth of the fill line respectively,
shown in yellow and red in Fig.~\ref{fig:setup}b.
The applied $\partial B_z/\partial z$ counteracts the intrinsic field inhomogeneity and thus
sharpens the bulk peak.}
\label{fig:imaging}       
\end{figure*}

The small $^3$He sample size and low filling factor, which necessarily arise from the experimental constraints on the sample cell discussed in the previous section and operation at microkelvin temperatures, impose severe demands of the required sensitivity of the NMR spectrometer. These have been met by a spectrometer based around a two stage DC SQUID with energy sensitivity 20h \cite{LevitinAPL}, Fig.~\ref{fig:spectrometer}. The NMR receiver coil (wound of copper wire onto the sample cell) forms part of a series tuned circuit with the SQUID input coil (the SQUID is located remotely at 1 K). This set-up achieves a noise temperature of 5 mK at the operating frequency of around 1 MHz. The NMR signal size can be reliably calculated (modulo knowledge of the NMR linewidth) using the principle of reciprocity.
The NMR static magnetic field is provided by a superconducting magnet located in the main $^4$He bath. The magnet is equipped with a field gradient coil and various shim coils. These allow us to perform simple one-dimensional NMR imaging to unambiguously identify components of the observed NMR signal. We can therefore distinguish beween the signal arising from liquid within the cavity and that arising from the small volume of bulk liquid near the fill line, Fig.~\ref{fig:imaging}. This bulk liquid marker is important, since it enables us to resolve the suppression of the $T_c$ of the superfluid film defined by the cavity and bulk liquid. 
\subsection{ Cooling the $^3$He film}
\label{sec:NF3}
A schematic diagram of the experimental set-up is shown in Fig.~\ref{fig:setup} . The cavity is supported by a structure fabricated from the machineable ceramic Macor. This cavity is located at the centre of the NMR magnet, and is mounted on a silver plate, which is cooled by the copper nuclear stage through a silver rod. This link is such that the temperature of the silver plate can be swept by a heater, without undue heat load on the cooling stage.

The $^3$He sample is thermalised by a silver heat exchanger mounted on the silver plate. The key feature is that the $^3$He in the cavity can be cooled to temperatures well below $T_c$ by conduction through the $^3$He column in the fill line. This exploits the high thermal conductivity of liquid $^3$He, which in the normal state exceeds that of copper at low mK temperatures. In this case the fill line is made of sterling silver, with an internal diameter of 0.57 mm. Before attempting this experiment there was concern about heat release from the nanofluidic cell materials, particularly
in view of the use of an amorphous material (glass) and potential stress release following the anodic bonding process. In an attempt to mitigate this risk, the upper and lower surfaces were silvered, and thermally anchored. In the event no serious time dependent heat leaks have been detected to date.

It is amusing to note that the cooling of this nanoscale sample of $^3$He in this way might be termed as ``cooling through the leads'', in analogy with the approach to cooling semiconductor nanostructures to ultralow temperatures, that is an important focus of Microkelvin.

\subsection{Cold valve}
\label{sec:NF4}
\begin{figure*}
  \includegraphics[width=1.0\textwidth]{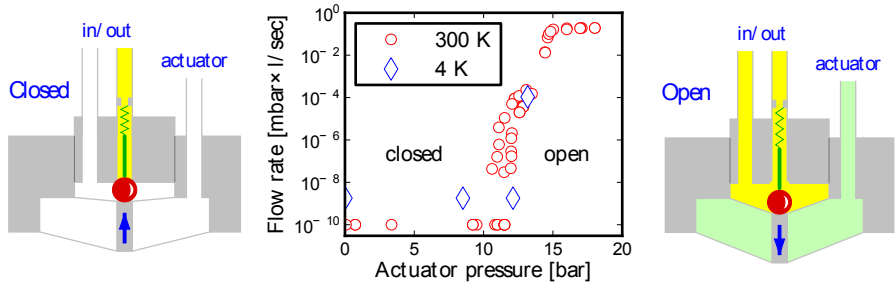}
\caption{Superleak-tight cryogenic valve.
The seal is between a stainless steel seat and a ruby ball.
The actuator with two flexible membranes pushes the ball into the seat,
unless pressurised to about 15 bar, where the valve opens.
This is a development of the design \cite{simplecoldvalve}
with a micro spring added to prevent the ball from sticking to the seat.
The resolution of the flow measurements was $10^{-10}$ and $10^{-9}$ mbar$\times$l$/$sec at room temperature
and 4~K respectively.}
\label{fig:coldvalve}       
\end{figure*}
As previously discussed, a key parameter of topological superfluidity in $^3$He films is the boundary condition for quasiparticle scattering. Our ability to study superfluidity in the thinnest films $D<\xi_0$ will depend on achieving fully specular surfaces to avoid suppression of $T_c$. There is a significant body of evidence that surface scattering can be tuned from diffuse to specular, by plating the cavity surface with a sufficiently thick superfluid $^4$He film. In the work performed so far we were unable to achieve fully specular conditions, as a result of the instability of a thick superfluid $^4$He film on the cavity surfaces in our arrangement. The proposed mechanism for this instability is discussed in \cite{Levitin}. In order to overcome this we have developed a cold valve, Fig.~\ref{fig:coldvalve} inserted in the cavity fill line, based on a previously reported scheme \cite{simplecoldvalve}, but with modifications to ensure reliable operation. The valve is leak tight to a superfluid $^4$He film. It is mounted at dilution refrigerator temperatures, and pressure-actuated using $^3$He.

\section{Superfluid $^3$He confined in a nanoscale geometry: Review of recent results }
\label{sec:R}

\begin{figure*}
  \includegraphics[width=1.0\textwidth]{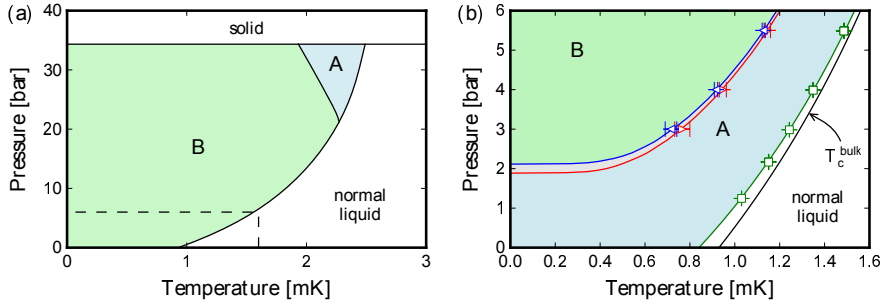}
\caption{Phase diagram of superfluid $^3$He in the bulk (a) and 0.7~um thick slab (b) \cite{Vollhardt,Levitin}.
Two superfluid phases are observed, A and B.
To stabilise the A phase in the bulk requires high pressure that enhances strong coupling.
The the slab the A phase is stabilised at low pressures due to confinement on a lengthscale comparable with superfluid coherence length. The superfluid transition in the slab is suppressed due to diffuse quasiparticle scattering at the cavity walls. Hysteresis between cooling and warming is observed at the A-B transition leading to the grey coexistence region.}
\label{fig:phasediagram}       
\end{figure*}

The main results of the experiments on a 0.7 \textmu{}m film \cite{Levitin,Levitin2013} are:

We observe the profound influence of confinement on the phase diagram, Fig.~\ref{fig:phasediagram}. The chiral A phase is stabilised at low pressure. This phase is identified by its negative NMR frequency shift (static field directed along cavity surface normal, corresponding to maximum in the dipole energy), confirmed by the tip angle dependence of this shift.

\begin{figure*}
  \includegraphics[width=1.0\textwidth]{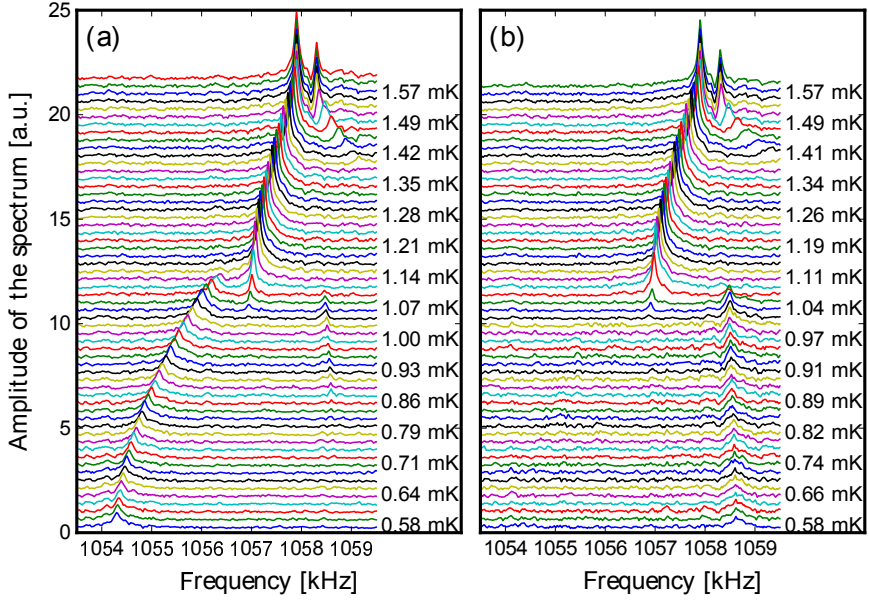}
\caption{Galleries of NMR spectra observed in a 0.7~\textmu{}m slab at 5.5~bar,
spectra offset vertically for clarity \cite{Levitin}.
The bulk marker is observed close to the superfluid transition temperature $T_c^{\mathrm{bulk}} = 1.5$~mK,
below which the marker is in the B phase with a positive frequency shift due to a wall-induced texture.
In the slab the A phase with a negative shift (dipole-unlocked texture) is stabilised by confinement between
the superfluid transition at $T_c^{\mathrm{slab}} = 0.98 T_c^{\mathrm{bulk}}$
and the AB transition at $T_{\mathrm{AB}} = 0.7 T_c^{\mathrm{bulk}} = 1.1$~mK,
Below $T_{\mathrm{AB}}$ two signatures of the B phase are observed stochastically,
B$_-$ with a negative shift larger than that in the A phase, and B$_+$ with a positive shift.
(a) Warm-up from the B$_-$ state; at $T=0.6 T_c^{\mathrm{bulk}} = 0.9$~mK
a small fraction of the sample undergoes an orientational transition into B$_+$ state,
which can be traced together with $B_-$ all the way up to $T_{\mathrm{AB}}$.
This transition is a manifestation of the metastable nature of the B$_-$ state,
which has higher dipole energy than B$_+$.
(b) Cool-down into B$_+$ state.}
\label{fig:ABportraits}       
\end{figure*}

At pressures above 2 bar a transition is observed to the planar-distorted B phase. We discovered that two inequivalent B-phase vacua can be stabilised, Fig.~\ref{fig:ABportraits}. These correspond to different relative orientations of the spin and orbit states of the pairs: the stable orientation, in which the projections of the spin and orbital angular momenta of the pairs are oriented according to $S_z=-L_z$, and a metastable state for which $S_z=L_z$. These two states are identified by distinct NMR frequency shifts, positive and negative, and by different tip angle dependences of the frequency shifts. Analysis allows us to determine two distinct averages of the planar distortion across the slab, which are in good agreement with the predictions of quasiclassical theory.

In some cases these two B-phase
states are observed to coexist, implying the existence of domain walls. Field cycling experiments identify these domain walls as textural, and suggest pinning by surface scratches. These topological defects are of interest for their potential to host majorana fermions \cite{VolovikIsing,Mizushima}; their controlled manipulation seems to be a possibility \cite{Levitin2013}.

The observed reduced temperature of the AB transition in the cavity, measured as a function of pressure up to 5.5 bar, is lower
than theoretical prediction. The enhanced stability of the A phase, at a seies of pressures, indicates that strong coupling effects play a role even at zero pressure. Here, on the other hand, bulk thermodynamic measurements, for example the heat capacity jump, take the weak coupling value.
Recent calculation in the weak coupling limit predicts the stablization by the cavity of a state with a spatially modulated order parameter which intervenes at the transition between the planar phase (degenerate with the A phase in the weak coupling limit) and the B phase \cite{VorontsovSauls}. No signature was observed of this putative striped phase. This motivates future work in thicker cavities, where the AB transition can be observed at zero pressure, as close as possible to the weak coupling limit; studying the influence of different surface boundary conditions is also of importance.

\begin{figure*}
  \includegraphics[width=1.0\textwidth]{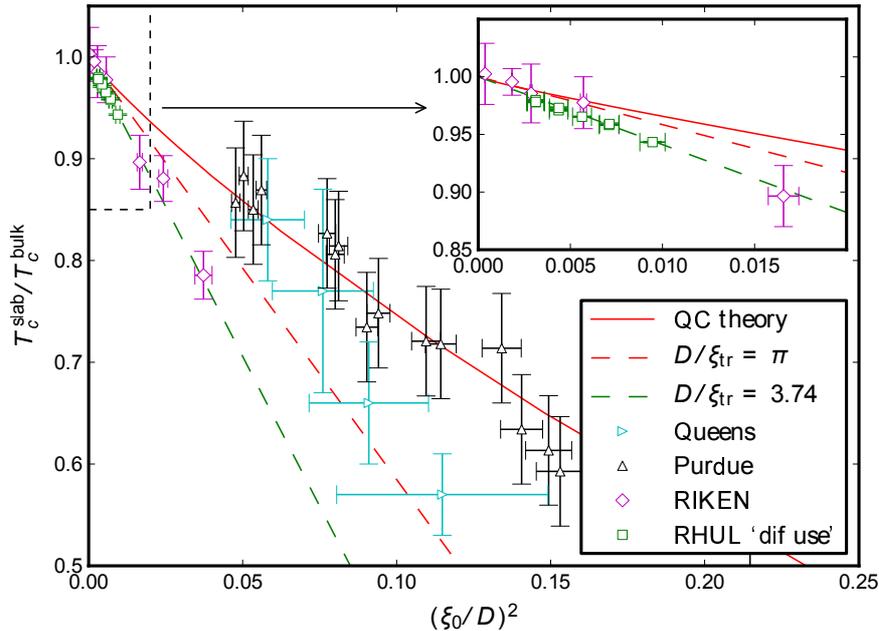}
\caption{The suppression of the superfluid transition temperature as a function of effective confinement. Data from RHUL \cite{Levitin}, Queens \cite{Steel}, Purdue\cite{Xu}, RIKEN\cite{Saitoh}. Inset shows blow up of RHUL and RIKEN data. Theoretical suppression is linear in chosen coordinates in Ginzburg-Landau regime. Prediction of quasiclassical(QC)theory \cite{Kjaldman} is also shown. Results from RHUL and RIKEN appear to show higher than predicted suppression. }
\label{fig:Tcsuppression}       
\end{figure*}

We were also able to measure and compare with theory the observed suppression of $T_c$, Fig.~\ref{fig:Tcsuppression}, exploiting the sharp frequency shift onset in both the cavity and the bulk marker as a precise indicator. In this case the film thickness $D$ is essentially constant, apart from the small distortion of cavity height with pressure; the coherence length, and hence the effective confinement, are tuned by pressure. This contrasts with measurements on saturated films, where the film thickness itself is varied at essentially zero pressure. The results for diffuse boundaries indicate a discrepancy with theory; $T_c$ appears to be more strongly suppressed
than expected. This motivates the extension of measurements to thinner cavities.

\section{Superfluid $^3$He confined in a nanoscale geometry: Summary, conclusions and future prospects}
\label{sec:conclusions}  
The experiments discussed here demonstrate that the study of superfluid $^3$He under controlled nanoscale confinement is feasible. The door is open to sculpture of the p-wave superfluid. The potential tuning parameters
are many: geometry; confinement length scale; surface profile; liquid pressure; magnetic field; flow; rotation.

So far we have focussed on flat slab geometries, of different cavity height. Small steps in cavity height might be used to stabilise domain walls (in conjunction with a magnetic field used to manipulate them). Larger steps in cavity height, or continuous variation in cavity height (eg parabolic) could be used to stabilize the A-B interface. Coupled with NMR we may attempt to identify the relics of brane-antibrane annihilation \cite{Bradley}. The surface profile can be tailored by nanolithography. As previously discussed we should be able to tune \emph{in situ} surface quasiparticle scattering from diffuse to specular by coating it with a $^4$He film. Pressure tunes the effective confinement \emph{in situ} through its influence on the superfluid coherence length. A weak oscillatory flow field in a torsional oscillator or Helmholtz resonator will determine the superfluid density. [In this work it is important to engineer a surface profile that will to effectively couple the normal component to the wall; this is based on the discovery of anomalous transverse momentum relaxation in normal state thin films \cite{Sharma}]. A stronger dc flow is predicted to induce strong order parameter distortions, which can be studied in combined NMR-flow experiments, leading ultimately to measurements of the critical velocity.

Confinement can lead to completely new order parameters such as the spatially modulated ``striped'' phase \cite{VorontsovSauls}. With increased confinement in a slab geometry (thinner films), finite size effects will play a role, and the spherical Fermi surface will break up into a series of Fermi discs. We may anticipate that the gapped chiral A-phase will be stabilized \cite{VolovikExotic}. However it is an open question whether confinement may rather stabilize the planar phase, which is degenerate with the chiral A-phase in the weak coupling limit, due to the effects of dimensionality of strong coupling effects of spin-fluctuation mediated pairing. Recent theoretical work has also demonstrated that the pairing symmetry is naturally influenced by placing the superfluid within a lithographically defined periodic structure (for appropriate small periodicities) stabilizing several new phases \cite{WimanSauls}.

The question arises: what are the stable topological defects of superfluid $^3$He under confinement? We have already demonstrated a new B-phase textural domain wall. We need to investigate whether the sought-after half quantum vortex (HQV) \cite{SalomaaVolovikHQV} is stable under some condition in rotating chiral A-phase under confinement. The HQV is technologically important because it hosts a majorana fermion at its core, and the braiding of such vortices can form the basis of topological quantum computing \cite{Nayak}.

Finally the study of confined planar-distorted B-phase is the ideal laboratory for the ``discovery'' of Majorana fermions, since the surface excitations form a gas of such particles. 

The prospect of adapting nanophysics techniques, and developing new ones, to perform experimental nanoscience on topological superfluids (both established and yet to be discovered), under different nano-confined conditions, is an exciting one. The purity of superfluid $^3$He and the complete absence of disorder, both unique to a condensed matter system, coupled with the high level of control we have over the surface and confinement parameters are key ingredients in this scientific methodology. In superfluid $^3$He we have a ``theory of everything'', which supports both the robust confrontation between theory and experiment, and the design of experiments to uncover the potential exotic responses of surface and edge excitations emerging from the quantum vacuum.

\begin{acknowledgements}
We thank S. Dimov, S. S. Verbridge for assistance with cell construction at Cornell; S. T. Boogert and A. Bosco for advice and loan of equipment for the optical thickness measurement. We thank G. Volovik, A. Vorontsov, J. Sauls and E. V. Surovtsev for theoretical input. This work was supported by: EPSRC grants EP/C522877/1, EP/E0541129/1, EP/J022004/1; NSF grants DMR-0806629, DMR-120991; European Microkelvin Consortium (FP7 grant agreement no. 228464).
\end{acknowledgements}



\end{document}